\documentclass[twoside,11pt,leqno]{article}

\advance\topmargin by -\headheight\advance\topmargin by -\headsep
\usepackage{amsmath}
\usepackage{amsthm}
\usepackage{amssymb}
\usepackage{amscd}
\usepackage{graphicx}
\usepackage{epsfig}
\numberwithin{equation}{section}

\theoremstyle{definition}

\numberwithin{equation}{section}

\allowdisplaybreaks
\begin{document}
\thispagestyle{empty}

\hfill{\small \textsl{Faddeev Memorial Volume, January 2018}}
\bigskip
\begin{center}
{\Large Remembering Ludwig Dmitrievich Faddeev \\
our lifelong partner in mathematical physics}
\footnote{\copyright{\footnotesize {2018 Daniel Sternheimer}}}
\end{center}

\begin{center}
{\large Daniel Sternheimer}
\\ \smallskip
\textsl{Department of Mathematics, Rikkyo University, Tokyo, Japan\\
\smallskip
\& Institut de Math\'ematiques de Bourgogne, Dijon, France, \\
\smallskip
Honorary Professor, St.Petersburg State University.}
\end{center}

\begin{abstract}
We briefly recount the long friendship that developed between Ludwig and 
us (Moshe Flato and I), since we first met at ICM 1966 in Moscow. That 
friendship extended to his school and family, and persists to this day. 
Its strong personal impact and main scientific components are sketched,
including reflexions on what mathematical physics is (or should be).    
 
\end{abstract}
\pagestyle{myheadings}

\section{In the beginning}

Since there are, and will certainly be (including in this volume) many
accounts of the seminal works of Ludwig Faddeev (23 March 1934 -- 26 February
2017) and their impact on modern science, especially modern mathematical 
physics and mathematical physicists, I shall concentrate this short 
contribution on events, sometimes anecdotical, that are characteristic 
and not so well known, and on reflexions on mathematical physics. 
Most involve my friend and coworker for almost 35 years, 
Moshe Flato (17 September 1937 -- 27 November 1998). 
Inevitably, these include also a number of major scientific developments 
and prominent Russian scientists. Though (at the start) younger than many of 
the latter, Ludwig quickly rose to their level. Like them (maybe even more 
than them) he created and developed his own school, a scienfific family, 
where appeared and flourished a number of leading mathematical physicists, 
now present in a variety of academic institutions in leading scientific 
countries. At the level attained, that was possible only in the USSR of that 
time, and only with a scientist with a vision. 

\subsection{ICM 1966 and around}

\subsubsection{First meeting and background}
 
Moshe Flato and I first met Ludwig Faddeev at the International Congress of 
Mathematicians that was held in Moscow (August 16-26, 1966), where the acronym 
ICM came into wide usage, and later that year when he spent two months at IHES
(cf. Ludwig's recollection in \cite{MFrec99}). It was ``love at first sight" 
between the young mathematical physicists that we were then (he was born in 
1934, Moshe in 1937 and I in 1938). Though we sometimes had to refer to him 
as ``Faddeev", between us we always called each other by first names. 
So for us he was always ``Ludwig" and not ``Ludwig Dmitrievich" (Russian 
style) or ``Professor Faddeev", a formal Western way to address colleagues. 

His first name is unusual for a Russian. As one may guess, it was given to him
in homage to Ludwig van Beethoven. Music (classical of course) was important in
his family, and for him as a corollary. In my mind I can still hear him humming
``in petto" some tune. Music was an interest that Ludwig shared with Moshe, 
who could play piano ``by ear" (from classical music to Gershwin) and at some 
point hesitated between a career in piano and in science. 

Ludwig decided to get higher education at the Department of Physics of 
Leningrad University, mostly to be independent of his father who was a 
renowned algebraist and a Professor at the Department of Mathematics. Yet 
the leopard can't change its spots (in French, ``chassez le naturel, il 
revient au galop") and Ludwig's starting point in research was more in 
mathematics, due to the influence of his tutor from the third year of 
undergraduate studies, Olga Alexandrovna Ladyzhenskaya, a renowned specialist 
in PDE, who however gave him full freedom to follow his own path. 

Similarly (see \cite{CMF99} for more details), after an undergraduate 
education in both mathematics and physics, and some hesitation between 
Shimshon Amitsur (ring theory), Nathan Rosen (relativity) and Giulio Racah 
(spectroscopy), Moshe settled for theoretical physics with Racah, 
starting by working in applications of group theory to physics. 
Another parallel is that a main reason that (in October 1963) brought 
Moshe to France, where he soon made a career in the mathematical community, 
was his desire to become independent of Racah, his ``scientific father".  

For different reasons, both did not choose between mathematics and physics, 
but became original mathematical physicists, at a time when that was 
not so fashionable (at least, in the way they pursued it). Thus they had many 
common points in spite of seemingly different backgrounds. Given their strong 
personalities, they could not remain indifferent to each other. As happened 
most of the time with these dedicated scientists, ``love at first sight" 
was the obvious outcome -- and naturally extended to me.           

Of course Moshe's chemistry with Russians owed a lot to his ``Russian genes",
including the fact he spoke Russian (and Polish, incidentally) very well, 
thanks to his parents and grandmothers. Moshe's musical ear was excellent, 
so his Russian accent was perfect, though Russians could not quite figure out 
from which Russian city he originated!  

\subsubsection{Mathematical Physics?}

An even more important factor for empathy was the fact that we were all 
(especially Ludwig and Moshe) true mathematical physicists, able (as Ludwig 
once told us) to ``speak mathematics" to mathematicians and ``speak physics" 
to physicists (the opposite is more common). 
Not many are truly bilingual in this way. As Sir Michael Atiyah said in 
London at the 2000 International Congress of Mathematical Physics \cite{Ati00} 
``mathematics and physics are two communities separated by a common language". 
The language is mathematics of course, but it is spoken with very different 
accents and grammars.  

Though the terminology may seem self explanatory, it is not. To be slightly
pedantic, one ought to distinguish between \textsl{mathematical physics}, i.e. 
(hopefully rigorous) mathematical treatment of physical problems, and   
\textsl{physical mathematics}, i.e. parts of mathematics inspired by physics.
[There is also a somewhat old-fashioned notion of mathematical physics, 
mainly in the UK, referring essentially to differential equations coming from
physics and engineering.] What is commonly called mathematical physics covers
in fact all these notions. 

Ludwig excelled in all aspects, and knew what he was doing. I remember a talk 
he gave in Moshe's seminar in Dijon in which (in part for pedagogical reasons) 
he went a bit too fast in his derivation. His co-author and son-in-law 
Sasha Volkov, a very honest and precise scientist who was then for a couple 
of years postdoc in Dijon, remarked during the talk that they had not yet 
proved a statement in that generality (it was eventually).
 
There is an important difference between rigorous mathematical physics and
pure mathematics. The purpose is often different. In mathematics one tends 
to study a problem in as general a context as possible (even if one does not
go as far as Bourbaki). While in mathematical physics, where the aim is usually
to tackle specific physical problems, and though generalizations may turn out 
to have unexpected consequences, it is often enough to use (and develop) 
mathematical tools adapted to the desired applications.   

The main difference between mathematical and theoretical physics is that 
mathematical rigor is a concern in the former but not in the latter. Some 
famous theoretical physicists do not quite understand why mathematical 
physicists are developing such an ingenious machinery to prove results 
they know for a long time to be true, and are often verified experimentally. 
Many others do not even realize that there may be a problem. The issue is not
purely academic, even for physicists, because it may happen that a rigorous
mathematical treatment points out to an effect that is real and potentially 
important.

Such distinctions developed with time. In the Antiquity there were practically 
none. Until a couple of centuries ago most leading scientists were both 
mathematicians and physicists, as is attested by the variety of theorems and 
laws that bear their name. Gradually science became a kind of Babel tower, 
even within a domain. One sometimes distinguishes between ``specialists", 
who know almost everything on a very restricted area, and ``generalists" who
know a lot (albeit often superficially) on many domains. [With humor, going 
to the limit, that can be expressed by saying that specialists know everything
on nothing while generalists know nothing on everything.] Ludwig was a rare
example of someone who was both, having a huge knowledge and being able to go 
deep in many domains.

\subsubsection{ICM 1966}
 
ICM 1966 was a major event for the development of mathematics. Until then
the scientific exchanges between USSR and ``the West" had been very limited. 
A record number of mathematicians attended (4282 according to official 
statistics), of which 1479 came from the USSR, 672 from other ``Socialist 
countries" in Europe, while over 1200 came from ``Western countries", 
including 280 from France: Moshe (then still only citizen of Israel but
working in France since October 1963) and I were among the latter. 
I remember that we and many from the French delegation traveled to Moscow
on a Tupolev plane, organized as a train with compartments seating eight. 

In Moscow we were accommodated, together with many ``ordinary" participants 
(some VIPs were accommodated in ``smaller" hotels closer to the Kremlin), 
in the huge hotel Ukraina (opened in 1957, the largest hotel in Europe),
one of the seven Stalinist skyscrapers in Moscow with a height of 206 meters 
(including the spire, 73 meters long) and total floor area ca. 88000 m$^2$, 
a small city in itself. On every floor there were ``etazhniks" supposed to help
but in fact checking on the guests. (We encountered the same system 10 years 
later in Taipei ...). 

The opening ceremony of ICM 1966, as it is now known, was held in the Kremlin. 
As far as I remember, it is during the long party which followed that
we met, and instantly became friends with, Ludwig and other leading Soviet
scientists like Israel Moiseevich Gelfand and Nikolay Nikolayevich Bogolyubov.
The latter (N.N.) who turned 57 during the Congress was then, inter alia, 
Director and main founder of the Joint Institute of Nuclear Sciences in Dubna 
(the ``Eastern CERN"), a new city which is now ``Science City Dubna". He invited 
us to spend a week there after ICM, which we did (that is another story), 
staying in Hotel Dubna, which was then a few months old; the building has not
changed since that time, but the rooms were of course eventually modernized.

N.N. was also Academician Secretary of the Division of Mathematics of the Soviet   
Academy of Sciences, a capacity in which Ludwig became his successor. N.N. was
not affiliated with the Division of Physics, mainly due to local scientific 
policy reasons. One of the factors was probably a conflict with Landau, who 
was a kind of "Russian Pauli" and in particular strongly objected Quantum Field 
Theory. In a centralized scientific community, several ``crocodiles" have a 
hard time to coexist in the same pond, however large it may be. The same
is true for France, especially in physics, and Moshe  \cite{CMF99}, like N.N. 
and Ludwig, quickly found ``scientific hospitality" in the more open 
mathematical community.
  
The bulk of the Congress sessions was located in another of the seven 
Stalinistic style skyscrapers, that of the Moscow State University.

\subsection{ICMP 1972 and some mathematical physics}

\subsubsection{ICMP 1972 and before}
The next occasion we could meet was, if I remember correctly, in Moscow in
December 1972, where the logo ``$M \cap \Phi$" was introduced. We may have met
also in Warsaw, earlier that month, in the ``Franco--Polish--Swedish" meeting 
on ``Fundamental problems in elementary particle physics", which evolved from
the Franco--Swedish ``Gunnar K\"all\'en Colloquia", initiated in 1967
by Moshe and Gunnar K\"all\'en (who, like Moshe, was not mincing his words) 
and so-named after K\"all\'en's accidental death at 42 in 1968. 
That may also have contributed to inspire Bogolyubov to organize 
``$M \cap \Phi$" just after our conference in Warsaw.  

After IAMP (the International Association of Mathematical Physics) was created 
in 1976, that large ``$M \cap \Phi$" meeting was a posteriori defined as the 
First International Congress of Mathematical Physics (ICMP). The following 
meeting in Warsaw in March 1974 (also an avatar of the Gunnar K\"all\'en 
colloquia), where we met again and where Moshe and I suggested the creation 
of IAMP (again that is another story, told in part by Rudolf Haag in 
\cite{RH10} and by Elliott Lieb in \cite{MFrec99}), was then defined as the 
Second ICMP, ICMP 1972 was organized by N.N. Bogolyubov. There was again a 
grand party in the Kremlin, during which Moshe had further extensive 
discussions with Ludwig and Bogolyubov. 

\subsubsection{Integrable systems, quantum groups and deformation quantization}

When we met with Ludwig for the first time, Ludwig and Moshe concurred that 
physics is based on a 4 dimensional space-time, so that two dimensional models 
were not true physics. But in 1970 a crucial ``phase transition" occurred in 
Ludwig's scientific interests \cite{FZ71}, albeit in a different scientific 
context. As he writes in \cite{ShawLF}:\\
``In 1970 I was introduced by V. Zakharov to the inverse scattering method of 
solving the nonlinear evolution equation on two dimensional space-time. 
Our first joint result -- the Hamiltonian interpretation and complete 
integrability of the Korteveg - de Vries equation -- defined my activity for 
20 years. The main achievements here, made together with a large group of 
excellent students (now called ``Leningrad School"), are the unravelling of 
the algebraic structure of quantum integrable models (the Yang-Baxter equation) 
and formulation of the Algebraic Bethe Ansatz. This development eventually 
became a base of construction of quantum groups by V. Drinfeld."

Incidentally, as he told us, the terminology ``Yang-Baxter" is due to Ludwig 
and Leon Takhtajan \cite{FT79} who showed that, implicit in Baxter's work, 
is the machinery which relates the eight-vertex model to a linear problem, and
gave a clear identification of its purely algebraic aspects. It seems that 
Baxter did not realize the paramount importance of the structure he had 
introduced, until Ludwig and Leon made it clear.   

Around the same time (in the 70s), building on Moshe's ideas about the role of 
deformations in physics (see e.g. \cite{dsWGMP32} and references therein) 
we started to develop what is now called ``deformation quantization". One of
the factors that led to ``the founding 1978 papers" \cite{BFFLS} was the fact 
that in 1975 at a meeting in Austin TX where I lectured on our works, 
Jerzy Pleba\'nski told me that it reminded him of Moyal, on whose work he had 
based lecture notes he had written (in Polish), a copy of which he eventually
sent us. I had not heard of Moyal's work, and on Mathematical Reviews what is 
now called the Moyal bracket was not even mentioned, but Moshe had heard of it 
so we ordered a copy from CNRS. Together with Jacques Vey's 1975 paper (who, 
inspired by our works on deformations of the Poisson brackets, rediscovered 
the Moyal bracket), and a lot of further study we made, that developed 
to \cite{BFFLS}.  

Interestingly, Ludwig told us much later that already in the late 60s he had 
lectured to his students in Leningrad on the passage from classical to quantum 
mechanics and on Weyl (and Moyal) quantization. That was also around the time 
(late 60s to mid 70s) when Felix Berezin was developing his approach to 
quantization (which roughly speaking he eventually defined as a functor from 
classical to quantum observables). But the lack of enough interactions in 
those days between various schools, especially on both sides of the 
``iron curtain", prevented what would have been a fruitful convergence 
to develop. Only a decade later, after quantum groups were introduced in 
Leningrad in the context of integrable systems, and the term coined by 
Drinfeld (together with the relation to our deformation quantization), 
it became clear that Ludwig and we were developing various aspects of 
essentially the same physical idea. 

And there is more, which surprisingly relates to our unconventional works 
in the 60s around symmetries of elementary particles, for which Moshe's 
approach followed from his education in the Racah school in Jeru\-salem 
(see a short account in \cite{dsWGMP32}). As Ludwig wrote in \cite{LF00}:\\
``An unfinished physical fundamental problem in a narrow sense is physics 
of elementary particles. This puts this part of physics into a special 
position. And it is here where modern mathematical physics has the most 
probable chances for a breakthrough."\\
That is in part what I am promoting lately \cite{dsWGMP32}, at least for
the symmetries part of it. Indeed (in a nutshell) I am advocating deforming 
the Poincar\'e group of special relativity, first to Anti de Sitter (with a 
tiny negative curvature), explaining the photons and leptons as composites 
(of ``singletons") and then for hadrons deforming it further to some quantum 
group version. Of course field theory needs to be built on that basis. 
We miss the impact Ludwig could give to such a project, helping solve the
``unfinished physical fundamental problem" he mentioned in 2000. 

\subsubsection{ICM 1990 in Kyoto and before}
In 1986-1990 Ludwig was President of the International Mathematical Union 
and then (it ended after him, but that is another story) ex-officio headed 
the Fields Medals committee for the Kyoto ICM 1990. In March 1990 we met him 
in London around the time of the final meeting of the Fields committee 
in Oxford. At some point he told us that he had a problem. He did not know
how to break the news to one of the laureates, whom incidentally we had known 
already when he was a graduate student. (In those days, email was not 
so widespread, especially when people were on the move.) Moshe told Ludwig: 
``No problem, in two days he is coming for dinner at my place in Paris." 

So Ludwig wrote the announcement, in his own very characteristic handwriting. 
Two days later when the guest came for dinner, Moshe congratulated him. 
He was sure Moshe was ``pulling his leg" but when he saw Ludwig's letter he 
realized that this was true and asked Moshe: ``Can I call my folks" and 
jumped on the phone in Moshe's office. He was one of the ``three quantum 
Fields medals" at ICM 1990 in Kyoto.
   
\section{Some scientific achievements and their recognition}

\subsection{Prizes and around} 
It is often said that the ways of the Lord are mysterious. So may seem to be 
the ways of the Nobel Committees and of many prizes committees, unless of 
course you are in the confidence, but then you are not supposed to tell. 

Moshe did know something (in great part indirectly, and I through him) 
because every year from 1971 to his death, since his independent and
original opinions were valued, he was asked (under Chapter~6 of the 
regulations) to nominate candidates for the Prize in physics. He also had 
``off the record" discussions with Swedish friends, of course not explicitely 
about the Prize. The last request to nominate arrived after his death in 
November 1998. Most of his nominations were handwritten by himself on the form 
he received, and some parts of these eventually made it through. It would be 
interesting for a historian of science, when these become available (not in 
our lifetime, which is the purpose of the cap of at least 50 years) to study 
his very original nominations in the light of what happened afterward. 

A few things may however be told, since they are in the public domain. 
One is that (at least for quite some time last century), accredited journalists 
who were waiting for the announcement of the Nobel prize were given three
closed envelopes containing the names of the laureates and the motivation
for the decision. When the announcement came, they were told to open one of
the three, and discard the other two without looking into these. [All did as
they were requested.]  So there must be at least twice more scientists who 
are of ``Nobel level" than there are actual laureates. [It can be reasonably
conjectured that, for different reasons, Ludwig and Bogolyubov belong to that
category.] A number of factors may tip the balance, including a part of luck 
as in any human context and extra-scientific factors, including living long 
enough for making room for a proper occasion. 

Several reasons make the Nobel prize more prominent than others. For a long      
time it was the prize carrying the highest award, though for every individual 
that depended on the number of laureates sharing it (at most 3, not always in 
equal parts) and for some time on how successful were the investments of the 
endowment. The main reasons are probably its ancientness and the grandiose 
Prize ceremonies (which Moshe attended a couple of times, including with me in 
1991 for its 90$^\mathrm{th}$ anniversary), that have no equivalent.  

Incidentally being awarded such a socially important recognition may become 
a curse for a scientist, in particular by encouraging complacency and in 
leading to that by many social duties. (Ludwig was spared that ``curse", 
and he remained active to his last day.) It is not easy for a Nobel laureate to 
continue and pursue research at a significant level, but there are exceptions. 
The Nobel medal has two sides, while that is in general not true for the 
Fields medal which must be awarded before the age of 40.   

In the Nobel 1999 physics prize announcements \cite{Nobel99}, one can see 
a filiation of ideas from QED (Quantum Electrodynamics) \cite{Nobel65} with its
Abelian gauge $U(1)$ to the 1954 construction by C. N. Yang and R. L. Mills 
of the first example of a nonabelian gauge theory, to the electroweak model 
introduced in the 60s by Weinberg, Salam and Glashow (in separate publications) 
which got the Nobel prize in 1979 after consequences were observed 
experimentally \cite{Nobel79} and to the works recognized in the 1999 Prize,
for which works by ``R. P. Feynman, B. S. DeWitt, L. D. Faddeev and V. N. Popov"
``made significant contributions." Indeed it was the 1967 paper and Kiev 
preprint of Faddeev and Popov which enabled 't~Hooft to prove in 1971 that the 
electroweak model is renormalizable, hence of physical relevance. 

There is a cute anecdote, which experts can understand. In December 1979, after
the Prize ceremony, a program was aired on Swedish radio, with Salam and 
Weinberg. At some point Weinberg quoted some phrase from the Bible. 
Salam (an immense scientist who well deserved the Nobel prize) remarked that 
it exists also in the Qur'an, to which Weinberg quipped: ``Yes, but we 
published it before!" 

One may wonder why Ludwig wrote such a short (2 pages) Physics Letters paper,   
very hard to understand, and gave the (important) details only in a Kiev 
preprint. There was a strict page limit in Physics Letters, and in Soviet Union 
in those days, longer papers could be sent for publication only after getting  
a permission from the Division of Nuclear Physics in the Academy, which was 
controlled by the Landau school who stood against Quantum Field Theory. Indeed
in those days \cite{LF00} ``Quantum Field Theory was virtually forbidden, 
especially in the Soviet Union, due to the influence of Landau." Still a 
number of scientists, who understood Russian, soon managed to get their hands 
on the Kiev preprint, and use its results. [The important Kiev preprint was 
translated into English only after 5 years, but preprints are not taken into 
consideration for the Nobel prize,.] 

More precisely the MSN review MR1773036 (2001j:81001) of \cite{LF00} reads:  
``The natural set of quantization rules for Yang-Mills fields, valid to all 
orders of perturbation theory, were obtained by L. D. Faddeev and V. N. Popov 
in 1966 and published in a short communication to Physics Letters 
[Phys. Lett. B 25 (1967), 29--30] and in an extended version as a preprint of 
the Kiev Institute of Theoretical Physics [``Perturbation theory for 
gauge-invariant fields'', Preprint No. FERMILAB-PUB-72-057-T/NAL-THY-57, 
Fermi Natl. Accel. Lab., Batavia, IL, 1972; translated from Preprint No. 
ITF-67-036, Inst. Theor. Phys., Acad. Sci. Ukraine, Kiev, 1967]." 

And the first part of Ludwig's citation for the 2008 Shaw prize (which 
carries a comparable amount) reads \cite{ShawLF}: ``Ludwig Faddeev has made 
many important contributions to quantum physics. Together with Victor Popov 
he showed the right way to quantize the famous Yang-Mills equations which 
underlie all contemporary work on sub-atomic physics. This led in particular 
to the work of 't~Hooft and Veltman which was recognized by the Nobel Prize 
for Physics of 1999." 

\subsection{Ludwig, an unusual group leader and great friend}

Victor Popov was one of Ludwig's first collaborators and, unlike most of his
collaborators, he was not technically his student since \cite{LF95} he started 
to work in physics with Yuri Novozhilov, with whom we became close friends 
when he spent 8 years at UNESCO in Paris. But Popov can be considered as the 
first member of his school, the famous ``Leningrad school" of mathematical 
physics.  

The second member was Ludwig's student Petr Petrovich Kulish (1944--2016), 
a good scientist and a fine person. His membership in the group was precious 
in many ways, both scientific and administrative. 

On the administrative part, a typical attitude of Ludwig towards the scientific
and political establishments is that he always was guided by the pursue of 
scientific excellence and, in case of problems, knew up to which point he 
could stretch the limit. 

Soon other members joined the group, who are still alive, many of whom must 
have contributed to this volume. I will give only one name because of an 
anecdote that is characteristic of the closely knit group it was. 

Indeed, after Alexander Rudolfowitsch Its (PhD 1977) got his Russian Doctor 
degree, there has been a party during our visit to Leningrad in 1989. 
The party was held in Its' small apartment, which essentially became one huge 
table. Though it was hard then to get good products, the many guests brought 
an impressive variety of goodies. That was probably the best party we ever 
had in Russia, in great part due to the exceptionally friendly and warm 
atmosphere. The material conditions eventually (and fortunately) improved in 
Russia, but the sense of solidarity that was felt then, typical of 
Ludwig's school, was unique.   

In addition to the above, the long overdue improvement of material conditions 
in Russia had another ``perverse effect" related to Ludwig and us. In Soviet 
times, Ludwig was one of the few who had a car. That permitted him to go to 
the forests near Leningrad and pick a large variety of excellent wild 
mushrooms of which (like many Russians) he was an expert, and which he would 
preserve in salt. Every time he was visiting us in Paris, he would bring a jar 
or two. After the demise of the USSR, ownership of cars was gradually 
democratized and Ludwig had to go further and further to find less and less 
mushrooms. To the point that a dozen years ago, when my late wife Noriko 
Sakurai and I visited his wife Anya and him in their house of Komarovo and 
went looking for mushrooms in the woods, Noriko was the only one who found 
some, small yellow chanterelles. At that time, when visiting Paris, Ludwig 
was instead bringing as presents excellent tea produced by his very 
successful son-in-law.

\subsection{Some other important works}

Ludwig had many important contributions, besides his seminal work with Popov. 
These are certainly detailed in the present volume, by persons more qualified 
than me to do so. And there are still some works ``in progress" as we say. 
I shall mention here only a few. 

As he wrote in his 2008 autobiography for the Shaw prize \cite{ShawLF}, 
his first scientific paper was published in 1956, so he has been involved in 
active scientific work for 60 years. ``[He] began by treating 
the mathematical questions of the quantum scattering theory, both direct 
and inverse problems. The treatment of the quantum scattering theory for the 
system of three particles, based on the integral equations, now bearing [his] 
name, brought [him his] first success. The work was highly appreciated by 
the specialists in nuclear physics. The attention of mathematicians came later 
and now the theory of many body quantum scattering is an active subject of 
modern mathematical physics." At some point it even seemed that this work 
might lead him to (at least part of) a Nobel prize in nuclear physics. 

In another part of his citation for the Shaw Prize (in 2008) one can read: 
``Faddeev also developed the quantum version of the beautiful theory of 
integrable systems in two dimensions which has important applications in 
solid state physics as well as in recent models of string theory." 
That is an area in which his ``Leningrad school" excelled, and one of the 
reasons why he moved to physics in two dimensions. That brought to his school 
very strong cooperation with important groups in Japan (especially the school 
of Mikio Sato in Kyoto) and in France (in addition to our strong interactions 
with him).    

The Shaw citation continues: ``In another application of the scattering 
theory of differential operators, Faddeev (jointly with Boris Pavlov) 
discovered a surprising link with number theory and the famous Riemann 
Hypothesis."

One of his more recent prophetic works can be found in his contribution to
the conference held in Dijon in 1999 after Moshe's untimely death in 
November 1998 (at the age of 61). In this apparently simple and elegant paper
\cite{LF99} he introduced ``The modular double [which] is just the double which 
defines the hidden symmetry in conformal field theory." 

I can conclude with one short sentence: Ludwig was a phenomenon.
 

\noindent 
       Daniel Sternheimer,
\vspace{1mm}

\noindent {Department of Mathematics, Rikkyo University, \\
\textit{3-34-1 Nishi-Ikebukuro, Toshima-ku, Tokyo 171-8501, Japan,}\\
\& Institut de Math\'ematiques de Bourgogne, Universit\'e de Bourgogne}\\
\textit {BP 47870, F-21078 Dijon Cedex, France.} \\ 
\texttt{Daniel.Sternheimer@u-bourgogne.fr}

\end{document}